# Non-Abelian inverse Anderson transitions

Weixuan Zhang[1*], Haiteng Wang[1*], Houjun Sun[2], and Xiangdong Zhang[1$]

[1]Key Laboratory of advanced optoelectronic quantum architecture and measurements of Ministry of Education, Beijing Key Laboratory of Nanophotonics & Ultrafine Optoelectronic Systems, School of Physics, Beijing Institute of Technology, 100081, Beijing, China

[2] Beijing Key Laboratory of Millimeter wave and Terahertz Techniques, School of Information and Electronics, Beijing Institute of Technology, Beijing 100081, China

*These authors contributed equally to this work.

$Author to whom any correspondence should be addressed. E-mail: zhangxd@bit.edu.cn

**ABSTRACT**

**Inverse Anderson transitions, where the flat-band localization is destroyed by disorder, have been wildly investigated in quantum and classical systems in the presence of Abelian gauge fields. Here, we report the first investigation on inverse Anderson transitions in the system with non-Abelian gauge fields. It is found that pseudospin-dependent localized and delocalized eigenstates coexist in the disordered non-Abelian Aharonov-Bohm cage, making inverse Anderson transitions depend on the relative phase of two internal pseudospins. Such an exotic phenomenon induced by the interplay between non-Abelian gauge fields and disorder has no Abelian analogy. Furthermore, we theoretically design and experimentally fabricate non-Abelian Aharonov-Bohm topolectrical circuits to observe the non-Abelian inverse Anderson transition. Through the direct measurements of frequency-dependent impedance responses and voltage dynamics, the pseudospin-dependent non-Abelian inverse Anderson transitions are observed. Our results establish the connection between inverse Anderson transitions and non-Abelian gauge fields, and thus comprise a new insight on the fundamental aspects of localization in disordered non-Abelian flat-band systems.**

Anderson localization [1] and flat-band localization [2] are two primary mechanisms to confine waves and the study of interplays between these two localization effects has aroused great interest [3-7]. Previous investigations have pointed out that the localization-delocalization transition can be induced by disorder in a lattice model where all energy bands are flatbands. Such a novel phenomenon is called inverse Anderson transition, which was originally predicted in the three-dimensional diamond lattice [4]. It has been pointed out that inverse Anderson transition is hard to be realized in low-dimensional systems, where the competition between geometric frustration and Anderson localization always exists. Contrary to such a wisdom, Longhi theoretically revealed that the antisymmetric-correlated disorder can trigger the appearance of inverse Anderson localizations in a quasi-one-dimensional (1D) Aharonov-Bohm (AB) cage [5]. This discovery greatly reduces the required dimension of inverse Anderson transitions, and promotes recent experimental realizations by electric circuits [6] and ultracold atoms [7]. To date, all studies on inverse Anderson transitions have been focused on systems with Abelian gauge fields.

Beyond Abelian gauge fields, the investigation of novel physics induced by non-Abelian gauge fields [8-22] has aroused great interest. Meanwhile, many novel phenomena related to non-Abelian physics have also been explored, such as the non-Abelian AB effect [17], non-Abelian band topology [23], and so on [24-27]. In particular, by generalizing magnetic flux-controlled destructive interferences of Abelian AB caging to the nilpotent interference matrix composed of non-Abelian gauge fields, the theoretical realization of non-Abelian AB caging has been proposed [28]. Different from Abelian AB caging, the spatial configuration of non-Abelian AB cage is sensitive to the nilpotent power of interference matrix and the initial state of the system. Although, those exotic properties of non-Abelian AB cage have been revealed, the study on the interplay between disorder effects and non-Abelian AB cage is still lacking. Hence, it is interesting to ask what novel phenomena can be revealed when non-Abelian gauge fields and inverse Anderson transitions meet together. And, how to observe non-Abelian inverse Anderson transitions in experiments.

In this work, we investigate the interplay between non-Abelian gauge fields and inverse Anderson transitions. We find that the pseudospin-dependent localized and delocalized eigenstates coexist in the disordered non-Abelian AB cage, making inverse Anderson transitions depend on the relative phase of two internal pseudospins. Furthermore, we fabricate non-Abelian AB topelectrical circuits to observe non-Abelian inverse Anderson transitions. Our work suggests a flexible platform to investigate the interaction among non-Abelian gauge fields, flat-band localizations and disorder, and may have potential applications in electronic signal control.

**The theory of non-Abelian inverse Anderson transitions.** We start to design a 1D flat-band model in the presence of a non-Abelian $U(2)$ gauge potential, as shown in Fig. 1a. There are three sublattices ('$a$', '$b$' and '$c$') in a single unit, and each sublattice contains two internal pseudospins ($|1>_{a,b,c} = [1, 0]^T$ and $|2>_{a,b,c} = [0, 1]^T$) with coupling matrixes being marked by $M_i$ ($i$=1, 2, 3 and 4). The Hamiltonian can be described as $H = \sum_{n=1}^{N}(M_1 a_n^+ b_{n-1} + M_2 a_n^+ b_n + M_3 a_n^+ c_{n-1} + M_4 a_n^+ c_n + h.c.) + U_a a_n^+ a_n + U_b b_n^+ b_n + U_c c_n^+ c_n$, where $\sigma_n^+ = [\sigma_{n,1}^+, \sigma_{n,2}^+]$ ($\sigma_n = [\sigma_{n,1}, \sigma_{n,2}]^T$) with $\sigma_n = a_n, b_n, c_n$ is the two-component creation (annihilation) operator with the internal pseudospin degree of freedom. $N$ is the total number of unit cells. $U_a$, $U_b$ and $U_c$ define on-site energies of three sublattices. The probability amplitude can be expressed as $|\psi> = \sum_{n=1}^{N} \sum_{\sigma}[\varphi_{\sigma_{n,1}}, \varphi_{\sigma_{n,2}}][\sigma_{n,1}^+, \sigma_{n,2}^+]^T |0>$, where $\varphi_{\sigma_{n,i}}$ ($\sigma = a, b, c$ and $i$=1, 2) corresponds to the $i$th pseudospin at '$\sigma$' sublattice in the $n$th unit.

It is worth noting that the non-Abelian AB caging can be realized based on the construction of nilpotent interference matrix [28]. In our model, the rightward-moving (leftward-moving) interference matrix is expressed as $I = 0.5(M_2 M_1 + M_4 M_3)$ ($I^\dagger$) with four coupling matrixes being set as $M_1 = M_4 = \begin{bmatrix} 1 & 0 \\ 0 & 1 \end{bmatrix}$, $M_2 = \begin{bmatrix} 1 & 0 \\ 0 & -1 \end{bmatrix}$, and $M_3 = \begin{bmatrix} 0 & -1 \\ 1 & 0 \end{bmatrix}$. In this case, the interference matrix possesses the nilpotent power of two ($I^2 = 0$). In addition, the non-Abelian gauge field requires that at least two hopping matrixes obeying the noncommutative relation, which is realized with $[M_2, M_3] \neq 0$. Based on the nilpotent interference matrix composed of non-Abelian hopping matrixes, the dual-path induced destructive interference of any initial states can appear in our model, corresponding to cases of $I(|1>_a + |2>_a) = 0$, $I^2(|1>_a - |2>_a) = 0$, and $I^{\dagger 2}(|1>_a + |2>_a = 0$, $I^\dagger(|1>_a - |2>_a) = 0$. In Fig. 1b, we numerically calculate non-Abelian Bloch bands. It is shown that complete flatbands appear at $E = \pm\sqrt{6}, \pm\sqrt{2}$ and $0$. The compact mode profiles of these non-Abelian flatbands are given in S1 of Supplementary Information.

Then, we introduce disorder into the non-Abelian AB caging. It has been pointed out that antisymmetric-correlated disorder can significantly break the destructive interference of $U(1)$ Abelian AB cages [5]. Such a property is still maintained in our $U(2)$ non-Abelian counterpart (See

S2 of Supplementary Information). Hence, we add antisymmetric-correlated disorder to onsite potentials at '*b*' and '*c*' sublattices in every odd unit, where onsite energies of '*b*' ($U_{b,n}$) and '*c*' ($U_{c,n}$) sublattices in the *n*th unit (*n* is odd) satisfy $U_{b,n} = -U_{c,n} = U_{dis,n}$ with $U_{dis,n}$ being an independent random number in the range of [-*W*, *W*], as shown in Fig. 1c. To quantify the influence of disorder on localization of non-Abelian AB cages, we calculate the variation of inverse participation ratios (*IPRs*) for eigenmodes around $E = \sqrt{6}$ as a function of the lattice length, as shown in Fig. 1d. Here, the disorder strength is set as *W*=0.25, and fifty disordered patterns are averaged. We can see that *IPRs* of nearly a half number of eigenmodes exhibit the 1/*N* scaling with low values. Meanwhile, high-valued *IPRs* for the other half of eigenmodes are nearly unchanged. These results indicate that localized and delocalized eigenmodes coexist in the disordered non-Abelian AB cage.

To further illustrate these localized and delocalized eigenmodes, we plot their spatial profiles in Figs. 1e and 1f. Bottom insets present enlarged views at '*a*' sublattices. As for the localized eigenmode, there are $\pi$ (0) phase differences between two pseudospins at '*a*' sublattices in even (odd) units, which are highlighted by green (orange) blocks. In contrast, phase differences of two pseudospins at '*a*' sublattices in even (odd) units equal to 0 ($\pi$) for the delocalized eigenmode. We can see that the relative phase of two pseudospins exhibit completely opposed features for localized and delocalized eigenmodes. In this case, two pseudospins can interference constructively or destructively under the influence of disorder on odd units, making localized and delocalized eigenstates coexist in the disordered non-Abelian AB cage. In S3 of Supplementary Information, spatial profiles of localized and delocalized eigenstates around other flat-band energies are presented. It is found that localized (delocalized) eigenstates always possess 0 ($\pi$) phase differences of two pseudospins at '*a*' sublattices in odd units. Different from localized eigenmodes around $E = \pm\sqrt{6}$ and 0, localized eigenmodes at $E = \pm\sqrt{2}$ possess vanished amplitudes at all disordered '*b*' and '*c*' sublattices, making its eigenspectra still exhibit the flat-band dispersion. In S4 of Supplementary Information, we also calculate variations of *IPRs* and eigenspectra with different disorder strengths. It is shown that localized and delocalized eigenmodes coexist with different disorder strengths, being consistent to density of states (DOSs) of disordered non-Abelian AB cages (see S5 of Supplementary Information).

Additionally, the coexistence of pseudospin-dependent localization and delocalization effects can be verified by wavefunction dynamics. Here, the initial excitation is added to two pseudospins at '*a*' sublattice in the 20th unit. As shown in Fig. 1g, the wavefunction is strongly localized with the initial phase difference being $\pi$ ( $|->_{a_{20th}} = |1>_{a_{20th}} - |2>_{a_{20th}}$ ). Nonetheless, the wavefunction extends to the whole structure with a zero-valued phase difference ($|+>_{a_{20th}} = |1>_{a_{20th}} + |2>_{a_{20th}}$ ), as shown in Fig. 1h. These results clearly manifest the existence of pseudospin-dependent non-Abelian inverse Anderson transitions.

Furthermore, to clarify the necessity of non-Abelian gauge fields for the coexistence of localized and delocalized eigenstates, we calculate wavefunction dynamics of disordered *U*(2) AB cages with Abelian coupling matrixes (see S6 of Supplementary Information). We find that pseudospin-independent localization or delocalization effect exists in disordered Abelian *U*(2) AB cages, indicating that the non-Abelian gauge field plays a key role for the mixture of pseudospin-dependent localized and delocalized eigenstates.

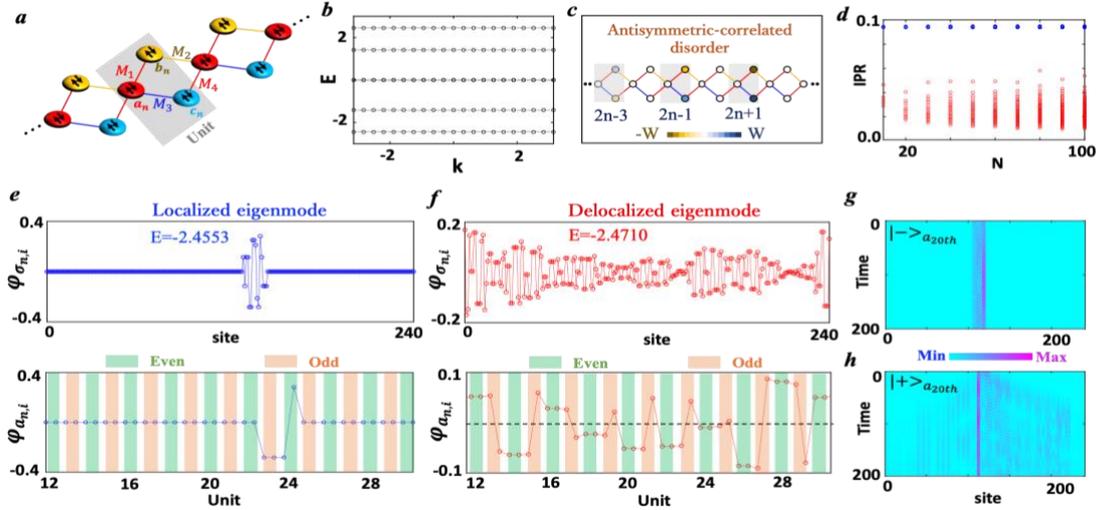

**FIG. 1**. **Theoretical results of non-Abelian inverse Anderson transitions.** (a). The non-Abelian flat-band model. (b). Flatbands of the clean non-Abelian AB caging. (c). The sketch map of the antisymmetric-correlated disorder on odd units. (d). The variation of *IPRs* for eigenmodes around E=$\sqrt{6}$ as a function of the lattice length with *W*=0.25. (e) and (f). Spatial profiles of localized and delocalized eigenmodes. (g) and (h). Wavefunction dynamics with initial phase differences between two pseudospins at '*a*' sublattice in the 20th unit equaling to $\pi$ and 0.

**Observation of non-Abelian inverse Anderson transitions in electric circuits.** Motivated by recent experimental breakthroughs in realizing quantum phases by electric circuits [29-62], we design electric circuits to observe the non-Abelian inverse Anderson transition. Figs. 2a and 2b display front and back sides for the unit of fabricated non-Abelian AB circuit. To realize the negative inter-site coupling, a pair of circuit nodes connected by the capacitor $C$ are considered to form an effective sublattice. Voltages at two nodes, which act as the $i$th ($i = 1, 2$) pseudospin of $\sigma$ sublattice ($\sigma = a, b, c$) in the $n$th unit, are expressed as $V_{1,\sigma_{n,i}}$ and $V_{2,\sigma_{n,i}}$. The coupling matrix can be realized by connecting nearest node pairs with capacitors $C$. Two left insets in Fig. 2c present circuit realizations of coupling paths mapped to $M_2M_1$ and $M_4M_3$. Each node is grounded by an inductor $L_g$, and the extra capacitor $2C$ is added for grounding on subnodes '$b$' and '$c$' to ensure the same resonance frequency. Three grounding capacitors $C_a$, $C_b$ and $C_c$ of '$a$', '$b$' and '$c$' sublattices are used to simulate onsite energies. To ensure the positive grounding capacitance on each node, a grounding capacitor $C_u = C$ is added to each node. Based on Kirchhoff equation, we can derive the eigenequation of our designed circuit, which is identical to that of the non-Abelian AB cage. See S8 of Supplementary Information for details. In particular, the probability amplitude $\varphi_{\sigma_{n,i}}$ is mapped to the voltage of $(V_{1,\sigma_{n,i}} - V_{2,\sigma_{n,i}})/\sqrt{2}$. The eigen-energy is directly related to the eigen-frequency of the circuit with $E = f_0^2/f^2 - 6 - C_u/C$. The onsite energy is given by $U_{a,b,c} = C_{a,b,c}/C$. Effective coupling matrixes possess the same form with that of non-Abelian AB cage.

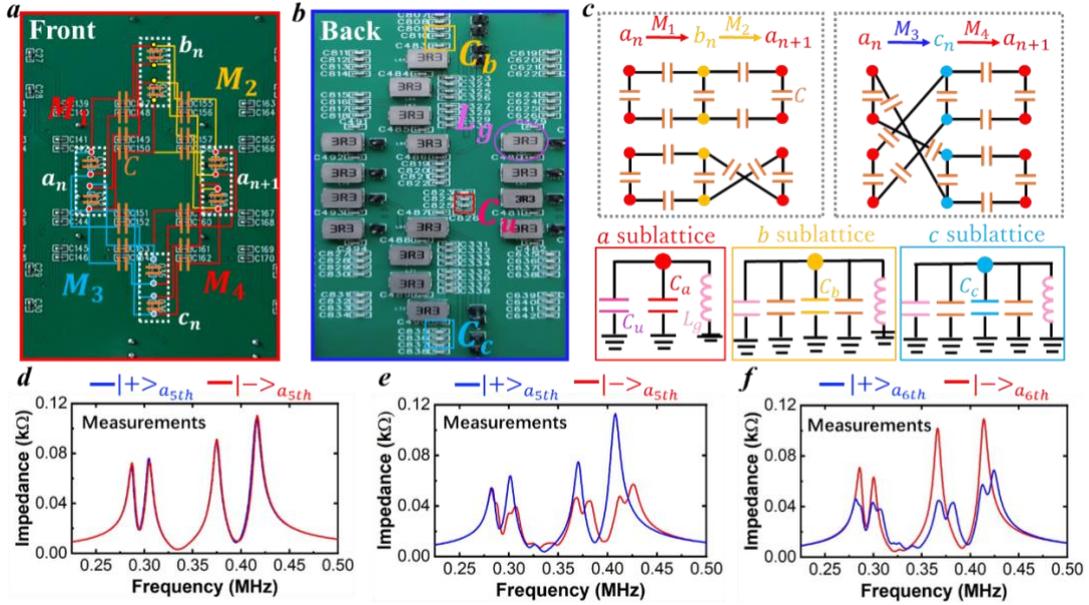

**FIG. 2**. **Observation of non-Abelian inverse Anderson transitions by impedance spectra.** (a) and (b). Photograph images of front and back sides of the circuit unit. (c) Diagrams of two moving paths $M_2M_1$ and $M_4M_3$ in the non-Abelian AB circuit. (d). Measured impedance spectra of '$a$' sublattice in the 5th unit of clean non-Abelian AB circuit. (e) and (f). Measured impedance spectra of '$a$' sublattices in 5th and 6th units of disordered non-Abelian AB circuit. Blue and red lines correspond to circuit excitations matching to $|+>_{a_{5th}}$ ($|+>_{a_{6th}}$) and $|->_{a_{5th}}$ ($|->_{a_{6th}}$).

At first, we focus on the clean non-Abelian AB circuit with $C$=10 nF, $L_g$=3.3 uH and $N$=10. It is well known that the impedance response of a circuit node is related to the local DOS of the mapped lattice model. We measure impedance spectra of the circuit node at '$a$' sublattice in the fifth unit, as shown in Fig. 2d. Blue and red lines correspond to results with excitations matching to $|+>_{a_{5th}}$ and $|->_{a_{5th}}$. It is shown that measured impedance spectra of two cases are identical. Specifically, there are four impedance peaks locating at 0.411 MHz, 0.371 MHz, 0.302 MHz and 0.285MHz, being matched to eigen-energies $E = -\sqrt{6}, -\sqrt{2}, \sqrt{2}, \sqrt{6}$ of the non-Abelian AB cage. The absence of impedance peak related to $E = 0$ is rooted in the spatial profile of zero-energy flatbands, where the probability amplitude at '$a$' sublattice is zero. These results clearly recover the pseudospin-degenerate flatbands in the clean non-Abelian AB cage.

Then, we turn to the non-Abelian AB circuit with disorder being added to '$b$' and '$c$' sublattices in odd units. Antisymmetric-correlated disorder can be easily introduced by setting the grounding capacitors $C_{b_n}$ and $C_{c_n}$ as random numbers in the range of $C$*[-0.25, 0.25] at different odd units. We note that non-Abelian inverse Anderson transitions can be characterized by the disorder-induced broadening of energy spectra and delocalization of eigenmodes. These phenomena correspond to the increased number and decreased value of impedance peaks in disordered non-Abelian AB circuit. To evaluate the pseudospin-dependent non-Abelian inverse Anderson transition, we measure impedance spectra of '$a$' subnodes in the fifth (sixth) unit, as shown in Fig. 2e (Fig. 2f). Blue and red lines correspond to results with circuit excitations matching to pseudospin-dependent states of $|+>_{a_{5th}}$ ($|+>_{a_{6th}}$) and $|->_{a_{5th}}$ ($|->_{a_{6th}}$), respectively. We can see that there are still four large-valued impedance peaks when the '$a$' sublattice in the odd (even) unit is excited in the form of $|+>$ ($|->$). This phenomenon is consistent to the theoretical prediction of localized eigenstates, which possess the 0 ($\pi$) phase difference between two pseudospins at '$a$' sublattice in odd (even) units. In addition, due to non-zero but very small amplitudes at '$a$' sublattices for localized eigenmodes around $E = 0$, there is also a little central impedance peak. Differently, it is found that lots of impedance peaks appear under the excitation matched to $|->$ ($|+>$) at '$a$' sublattice in the odd (even) unit. The increased number of impedance peaks clearly illustrates the broadening of pseudospin-dependent eigenspectrum. In addition, the value of impedance peaks is also decreased, being consistent to the delocalization effect. These results are consistent with DOSs of disordered non-Abelian AB cages, and can prove the coexistence of pseudospin-dependent localized and delocalized eigenstates. Simulation results of impedance responses are given in S9 of Supplementary Information. A good consistence between simulations and measurements is obtained.

It is found that the larger the circuit loss is, the wider of impedance peaks with a reduced number become. While, the impedance responses of lossy circuits can still reflect key properties of pseudospin-dependent non-Abelian inverse Anderson transitions.

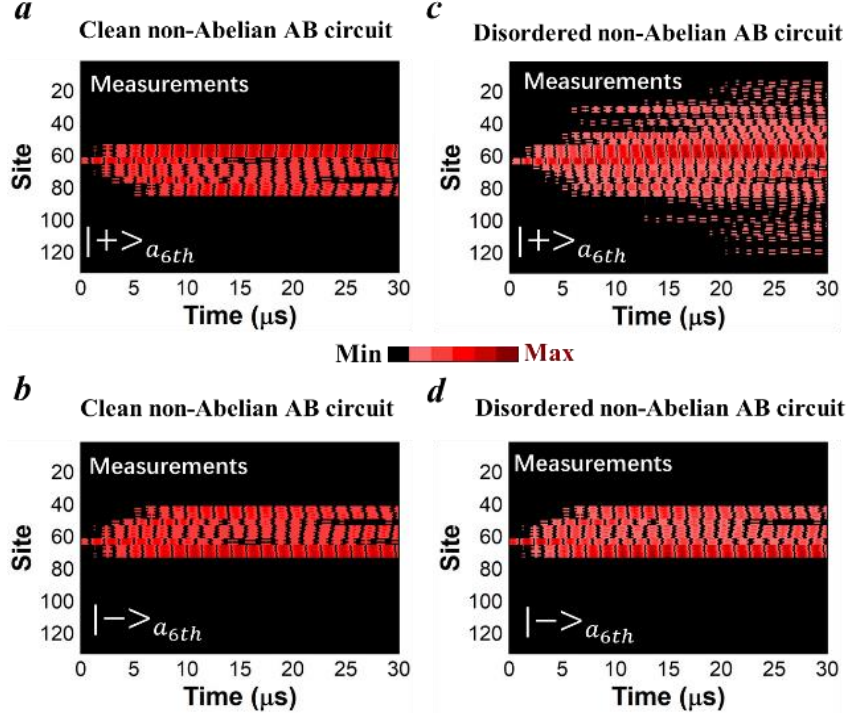

**FIG. 3. Observation of non-Abelian inverse Anderson transitions by voltage dynamics.** (a) and (b). Experimental results of voltage dynamics for the clean non-Abelian AB circuit with excitations matching to two-pseudospin states of $|+>_{a_{6th}}$ and $|->_{a_{6th}}$. (c) and (d). Voltage dynamics of disordered non-Abelian AB circuits with excitations matching to $|+>_{a_{6th}}$ and $|->_{a_{6th}}$.

Except for the measurement of impedance spectra, non-Abelian inverse Anderson transitions can also be directly observed with localization and delocalization effects of voltage dynamics. In this case, we measure voltage dynamics with different excitations. In particular, the first type of excitation $|+>_{a_{6th}}$ is in the form of $V_{1,a_{6,1}} = V_0 e^{iw_e t}, V_{2,a_{6,1}} = -V_0 e^{iw_e t}, V_{1,a_{6,2}} = V_0 e^{iw_e t}$, $V_{2,a_{6,2}} = -V_0 e^{iw_e t}$. The second one is defined by $V_{1,a_{6,1}} = V_0 e^{iw_e t}, V_{2,a_{6,1}} = -V_0 e^{iw_e t}, V_{1,a_{6,2}} = -V_0 e^{iw_e t}, V_{2,a_{6,2}} = V_0 e^{iw_e t}$, being consistent to the eigenmode of $|->_{a_{6th}}$. The associated excitation frequency is set as $w_e/2\pi = 0.33$ MHz (matched to E=0).

We firstly consider the clean non-Abelian AB circuit. Figs. 3a and 3b present experimental results of voltage dynamics under above two types of excitations. It is clearly shown that voltage signals all concentrate around the input node. This phenomenon clearly illustrates the flat-band localization of the non-Abelian AB circuit. In addition, we also note that localization domains are different under two types of excitations. This is due to the asymmetric outputs of right- and leftward interference matrixes $I$ and $I^\dagger$ acting on two pseudospin-dependent states.

Then, we measure the voltage dynamics in the disordered non-Abelian AB circuit. It is clearly shown that the fast extension appears with the excitation matching to $|+>_{a_{6th}}$, as shown in Fig. 3c. This phenomenon clearly illustrates the delocalization effect in the disordered non-Abelian AB cage. While, based on the same disordered AB circuit, the input voltage exhibits a significant localization effect when the initial excitation is in the form of $|->_{a_{6th}}$, as shown in Fig. 3d. These experimental results are consistent with simulations (in S10 of Supplementary Information). In this case, the experimental voltage dynamics clearly illustrates the key property of non-Abelian inverse Anderson transitions, where localization and delocalization effects coexist in the disordered non-Abelian AB cage.

In conclusion, we have revealed an exotic phenomenon of pseudospin-dependent non-Abelian inverse Anderson transitions. It is found that localized and delocalized eigenstates coexist in the non-Abelian flat-band system with disorder, and the disorder-induced localization-delocalization transition depends on the relative phase of two pseudospins. In experiments, we have fabricated non-Abelian AB topolectrical circuits to observe non-Abelian inverse Anderson transitions. Our work suggests a flexible platform to investigate the novel physics of disordered non-Abelian flat-band systems, and can enrich our understanding of disorder-induced localization-delocalization transitions.

**ACKNOWLEDGMENTS**.

This work was supported by the National Key R & D Program of China under No. 2022YFA1404900 and the National Natural Science Foundation of China under No.12104041.